\documentclass[10pt,preprint,a4paper]{aastex}
\usepackage{amsmath}                % American Mathematical Society package
\usepackage{amsfonts}               % American Mathematical Society fonts
\usepackage{amssymb}                % American Mathematical Society symbol
\usepackage{epsfig}                 % EPS figures
\usepackage{float}
\usepackage{graphicx}

\newcommand{\s}{{~\rm s}}

\newcommand{\K}{{~\rm K}}
\newcommand{\erg}{{~\rm erg}}
\newcommand{\yr}{{~\rm yr}}

\newcommand{\kpc}{{~\rm kpc}}

\begin{document}

\title{What planetary nebulae can tell us about jets in core collapse supernovae}

\author{Ealeal Bear\altaffilmark{1} and Noam Soker\altaffilmark{1}}

\altaffiltext{1}{Department of Physics, Technion -- Israel Institute of Technology, Haifa
32000, Israel; ealeal@physics.technion.ac.il, soker@physics.technion.ac.il}

\begin{abstract}
We compare the morphology of the core collapse supernova remnant (CCSNR) W49B with the morphology of many planetary nebulae (PNe), and deduce the orientation of the jets that shaped this CCSNR and estimate their energy. We find morphological features that are shared by some PNe and by the CCSNR W49B. In PNe these features, such as a barrel-shaped main body, are thought to be shaped by jets. We use these morphological similarities to deduce that the jets that shaped SNR W49B were launched along the symmetry axis of the `barrel', and to speculate that this CCSNR has two opposite lobes (or ears), that are too faint to be observed. We crudely estimate that the jets that shaped the CCSNR into a barrel shape had a kinetic energy that amounts to about one quarter to one third of the energy of the entire CCSNR. The morphological similarities strengthen the suggestion that jets play a central role in the explosion of massive stars.
\end{abstract}

Stars: jets - (ISM:) planetary nebulae: general - 
(stars:) supernovae: individual: SNR W49B

% ==========================================================
\section{INTRODUCTION}
\label{sec:intro}
% ==========================================================
 
Observations in recent years (e.g., \citealt{Maundetal2007, Lopezetal2011, Lopezetal2013, Lopezetal2014, Milisavljevic2013, Gonzalezetal2014, FesenMilisavljevic2016, Inserraetal2016}) suggest that jets play a role in many core collapse supernovae (CCSNe).
Long gamma ray bursts (GRBs) that are thought to be emitted by jets (e.g., \citealt{Woosley1993, ShavivDar1995, SariPiran1997}), and in some cases are accompanied by Type Ic supernovae (e.g., \citealt{Canoetal2016}), further support the notion that jets play a role in many CCSNe.

Several arguments and numerical simulations suggest that the collapse of a pre-explosion rapidly rotating core leads to the formation of two opposite well collimated jets (e.g. \citealt{Khokhlovetal1999, MacFadyen2001, Hoflich2001, Woosley2005, Burrows2007, Couch2009, Couch2011, TakiwakiKotake2011, Lazzati2012, Mostaetal2014, BrombergTchekhovskoy2016, Gilkis2017, Nishimuraetal2017}).
The condition of a rapidly rotating core requires that a stellar binary companion enters the envelope and spirals-in to the core. Therefore, pre-explosion rapidly rotating cores are rare.
In might be that jets in CCSNe are more common than the constraints imposed by this condition (e.g. \citealt{GrichenerSoker2017}).

It is possible that most, or even all, CCSNe are exploded by jets in what is termed the jet feedback mechanism (JFM; \citealt{Gilkisetal2016, Soker2016}). When there is a large amount of pre-collapse angular momentum, the mass accreted on to the newly born neutron star or black hole forms an accretion disk that launches opposite jets in a well defined axis, as described in the studies cited above.
When the pre-collapse core rotates slowly, the angular momentum of the accreted gas is likely to be stochastic \citep{GilkisSoker2014, GilkisSoker2015}, and intermittent accretion disks or belts might form \citep{SchreierSoker2016}, and might launch jets with stochastically varying directions; this is termed the jittering jets mechanism (\citealt{PapishSoker2011, PapishSoker2014a, PapishSoker2014b}).

\cite{Castellettietal2006} suggested that the CCSN remnant (CCSNR) Puppis A with a morphology that contains two opposite `ears', was shaped by jets. \cite{GrichenerSoker2017} extended the study to several other CCSNRs with ears. Under the assumption that the ears were shaped by jets that where launched during the explosion, \cite{GrichenerSoker2017} calculated the energy of the jets. The typical energy of the shaping jets is $\approx 10 \%$ of the kinetic energy of the CCSNR.
This is compatible with the last jet-launching episode in the jittering jets model.

Ears are observed also in supernova remnants (SNR) of Type Ia supernovae (SNR Ia).
These were compared to the morphologies of ears in planetary nebulae (PNe), and lead to the conclusion that some SN Ia explode inside PNe (term SNIP; \citealt{TsebrenkoSoker4472015, TsebrenkoSoker4502015}), while others might be shaped by jets launched during the explosion \citep{TsebrenkoSoker2013}.
Ears, and similar smaller and larger opposite lobes, in PNe are thought to be shaped by jets (e.g.,  \citealt{Soker1990AJ, SahaiTrauger1998}).

Jets are not the only agent that shape PNe (e.g., review by \citealt{BalickFrank2002} and \citealt{Demarco2014}). For example, a strong binary interaction can lead to a dense equatorial outflow that forms an equatorial ring, and might also collimate the fast wind later blown by the central star. For example, the simulations conducted by \cite{HuarteEspinosa2012} who included asymmetrical wind from the asymptotic giant branch (AGB) stellar progenitor, jets, a fast spherical wind from the central star, and ionization by the central star. The strongest support to the formation of lobes (ears) by jets is that in many PNe we observe multiple pairs of lobes (e.g. \citealt{SahaiTrauger1998, Sahai2000}). Precession of the disk that launches the jets naturally accounts for multipolar structure. Single-star models cannot account for multipolar morphologies.

We do note that in most PNe jets are not observe directly. The jets are active for a relatively short time before the PN phase. Examples for PNe where jets are observed are Abell~63 \citep{Mitchelletal2007}  and Fleming~1 \citep{Boffinetal2012}. Both these PNe have central binary systems, and in Fleming~1 the jets precess. In any case, when we refer to jets as the shaping agent of lobes, one should keep in mind that other shaping processes might take place, like binary ejection of dense equatorial gas.

In this study we examine the morphology of CCSNR W49B (section \ref{sec:W49B}) and compare it with the morphologies of several PNe (section \ref {sec:PNe}). In section \ref{sec:jets} we estimate the energy of jets that shaped the SNR. Like \cite{Lopezetal2011} and \cite{Lopezetal2013} we conclude that the CCSN was driven by jets, but we find the jets' direction to be perpendicular to the direction inferred by them \citep{Gonzalezetal2014}.
We summarize in section \ref{sec:Summary}.

% ==========================================================
\section{THE SNR W49B}
\label{sec:W49B}
% ==========================================================
W49B is one of the brightest supernova remnant (SNR) in X-rays in the Galaxy with $L_x \approx 10^{38} \erg \s^{-1}$ (e.g., \citealt{Lopezetal2013} and references within). The \cite{HESS2016} detect a $\gamma$-ray source coincident with the SNR W49B.
Some properties of the SNR W49B are summarized in Table 1. % \label{Tab:Table1
% TTTTTTTTTTTTTTTTTTTTTTTTTTTTTTTTTTTTTTTTTTTTTTTTTTTTTTTTTTTTTTTTTTTTTTTT
\begin{table}[H]
\label{Tab:Table1}
\begin{center}
\begin{tabular}{|c|c|c|}
  \hline
  Property & Value & Reference \\
  \hline
  Age & 1000 - 4000 \yr & A \\
  \hline
  Distance  & $\sim 8-10\kpc$ & B \\
  \hline
  Progenitor mass   & $\simeq 25M_\odot$ & C \\
  \hline
  Hot dust $(151 \pm 20\K)$ & $7.5 \pm 6.6\times 10^{-4} M_\odot$ & D \\
  \hline
  Cold dust $(45 \pm 4\K)$ & $6.4 \pm 3.2 M_\odot$ & D \\
  \hline
\end{tabular}
\end{center}
 \caption{Some properties of the SNR W49B. References are as follows. A: e.g., \cite{HESS2016, Micelietal2006} and references within.
B: \cite{MoffettReynolds1994,Zhuetal2014} and references within the latter.
C: An estimate zero-age main sequence mass of the progenitor (e.g., \citealt{MaedaNomoto2003, Nomotoetal2006, Gonzalezetal2014, Zhuetal2014}).
D: \cite{Zhuetal2014}. }
\end{table}
% TTTTTTTTTTTTTTTTTTTTTTTTTTTTTTTTTTTTTTTTTTTTTTTTTTTTTTTTTTTTTTTTTTTTTTTT

Previous studies of SNR W49B show three morphological and spectral features that are not found in other Galactic SNRs.
\begin{enumerate}
  \item Segregation of its nucleosynthetic products. The X-ray emission is elongated in a bar-like centrally shaped structure, with two plumes and is not homogeneously spread. In fact, the iron is missing in the west side while other elements (like silicon and sulfur) are relatively homogeneously distributed (for more details see \citealt{Hwangetal2000, Micelietal2006, Lopezetal2009a, Lopezetal2013}).
  \item The X-ray spectrum indicates rapid cooling. Rapid cooling is uncommon in young SNR such as W49B (for more details see \citealt{Ozawaetal2009,Lopezetal2013} and references within the latter).
  \item Detected X-ray spectral lines from chromium and manganese that are products of silicon burning(for more details see \citealt{Hwangetal2000, Micelietal2006, Lopezetal2013}).
\end{enumerate}

There are two proposed scenarios to account for the properties of SNR W49B, namely, a non spherical explosion of a massive star and an interaction of a spherical SN with an inhomogeneous ISM (e.g., \citealt{Lopezetal2013, Zhuetal2014}).
Recent studies support the first scenario of a jet-driven CCSN explosion over the ISM scenario \citep{Lopezetal2013, Gonzalezetal2014}.

\cite{Keohaneetal2007} suggest four distinguishing characteristics of a remnant of a jet-driven explosion of a massive star. (1) A double `T-shaped` structure. This structure follows the path of the bipolar jets. (2) A  non homogeneous spread of heavy elements which includes higher abundance of heavy elements than a typical Type II SN due to the fact that the jets originate from inside the iron core. (3) A supermassive progenitor with strong stellar winds. (4) No trace/observation of a neutron star (NS).

According to \cite{Keohaneetal2007} all of them are evident for SNR W49B. \cite{Zhuetal2014} also mentioned similar morphology features that support the explosion scenario.
The explosion scenario may account for inhomogeneous distribution of iron due to the fact that heavy elements tend to be ejected along the polar axis. \cite{Khokhlovetal1999} who numerically studied the explosion of supernovae caused by supersonic jets showed that the end result is a non-spherical supernova explosion with two polar directed jets. The composition of the jets must reflect the composition of the innermost parts of the star and should contain heavy and intermediate-mass elements.
The interaction with the interstellar medium (ISM) is relevant for other sources (e.g., G292.0+1.8: \citealt{Parketal2004}) but it requires fine-tuning of the configuration of the cloud \citep{Keohaneetal2007}.
In the present study we compare the morphology of SNR W49B with those of some PNe. The comparison of composition is not relevant to our comparison, as the nucleosynthesis in PNe and CCSN are very different from each other. We will therefore not mention the distribution of elements when comparing SNR W49B with PNe. 

\begin{figure}
\centering
\includegraphics[scale=0.85]{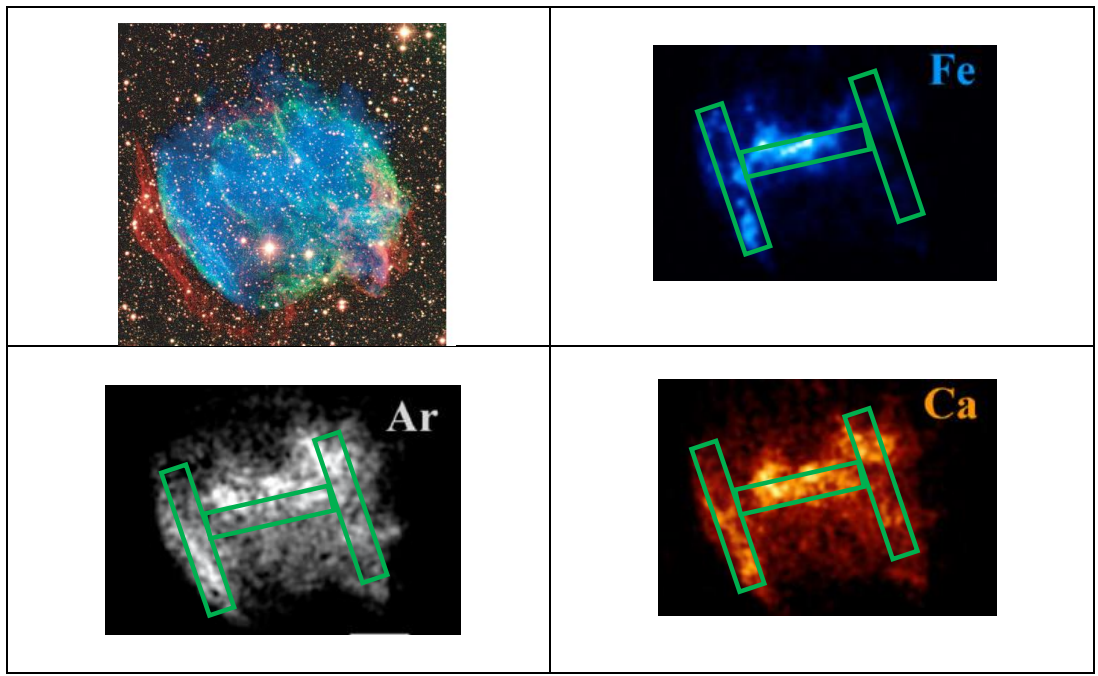}
\vskip -12.5 cm
\caption{The morphology of SNR W49B (upper left; from Chandra website: http://chandra.harvard.edu/photo/2013/w49b/), and the `H-shaped` morphology in X-ray lines of Fe~xxv (upper right), Ar~xvii and Ar~xviii (lower left), and Ca~xix and Ca~xx (lower right); the later three images are from \cite{Lopezetal2013}.
 The `H-shaped` morphology is marked in green.   }
\label{fig:SN_H_shaped}
\end{figure}

The double `T-shaped' structure mentioned by \cite{Keohaneetal2007} is addressed by us as the `H-shaped' morphology due to the fact that the legs of the `H' are not symmetrical and can form either a `H' shape (if both legs are observed) or a `T' shape (if only one leg is observed). This is seen in Fig. \ref{fig:SN_H_shaped}  where the two legs of the `H' do not have the same size, and therefore they can account for the T shape feature. We show the versatility of the asymmetric `H-shaped' of SNR W49B in Fig. \ref{fig:SN_H_shaped} that presents X-ray images of three elements.
% FFFFFFFFFFFFFFFFFFFFFFFFFFFFFFFFFFFFFFFFFFFFFFFF
% FFFFFFFFFFFFFFFFFFFFFFFFFFFFFFFFFFFFFFFFFFFFFFFF

Based on the morphologies of many PNe that are believed to have been shaped by jets, we do not accept all four properties of SNRs formed by jet-driven CCSN as listed by \cite{Keohaneetal2007}. We do not think, for example, that H or T shapes are the main outcomes of the operation of jets. We rather adopt the view, which is based, among others, on properties of PNe, that the most prominent feature of strong jets is the presence of two opposite ears or lobes (e.g., \citealt{GrichenerSoker2017}). We further adopt the view that the non-negligible amount of angular momentum of the exploding star might leave behind a rotating neutron star that is a pulsar or a magnetar, as seen in some CCSNRs with two opposite small lobes (termed ears) that are thought to be shaped by jets \citep{GrichenerSoker2017}.

% ==========================================================
\section{COMPARISON WITH PLANETARY NEBULAE}
\label{sec:PNe}
% ==========================================================

\begin{figure}[htb]
\centering
\includegraphics[scale=0.8]{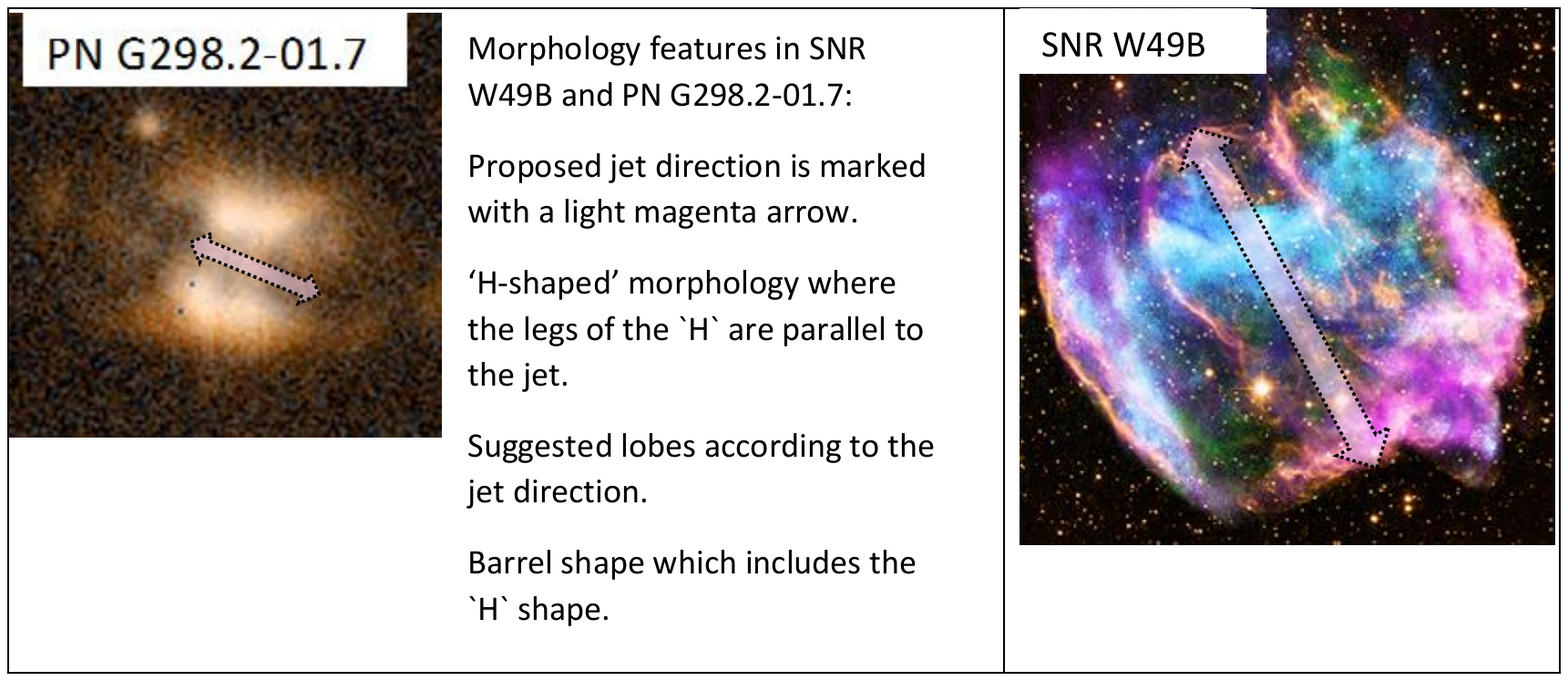}
\vskip -10.5 cm
\caption{Comparing the morphologies of the CCSNR W49B (credit the Chandra website http://chandra.harvard.edu/photo/2013/w49b/) and PN~G298.2-01.7 (He 2-76; credit \citealt{Balick1987,Gornyetal1999}). Comparing the morphology of PN~G298.2-01.7 with the SNR W49B.
Proposed direction of the axis of the jets in each case is marked with a light magenta double-head arrow. }
\label{fig:1PN_elaborated}
 \end{figure}

\begin{figure}[htb]
\centering
\includegraphics[scale=0.7]{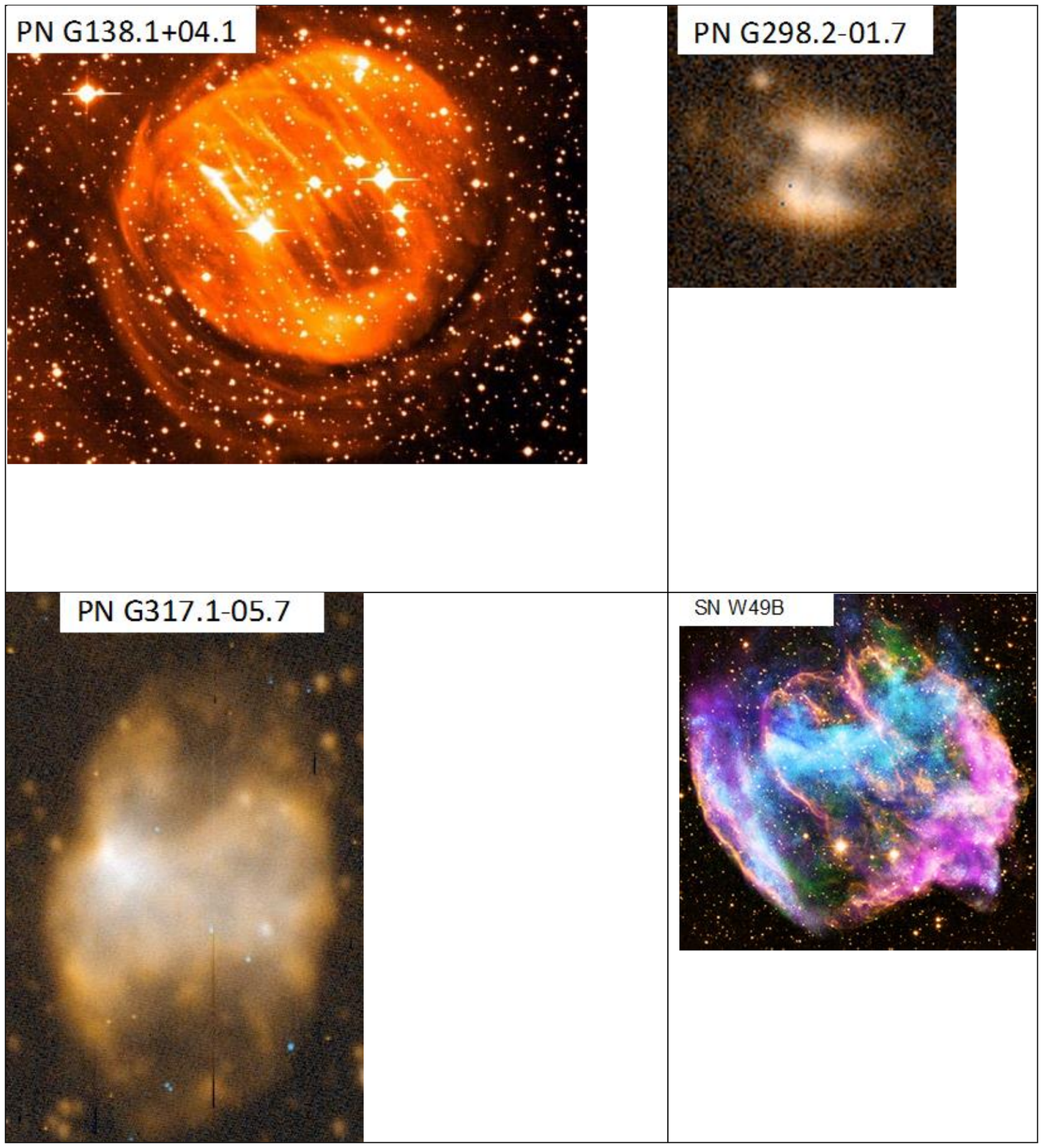}
\vskip -2.5 cm
\caption{Comparing the morphologies of the CCSNR W49B (credit the Chandra website http://chandra.harvard.edu/photo/2013/w49b/) and three PNe. PN~G138.1+04.1 (Sh 2-200; credit \citealt{Balick1987,Corradietal2003}) which contains many dense thin filaments,
PN~G298.2-01.7 (He 2-76; credit \citealt{Balick1987,Gornyetal1999}) and PN~G317.1-05.7 (He 2-119; credit \citealt{Balick1987,Gornyetal1999}).}
\label{fig:3PN_NOFEATURES}
 \end{figure}

The SNR W49B has a unique morphology among CCSNRs, but a morphology that is common among PNe, in  particular these four features. (1) The general barrel shaped of SNR W49B is observed in the central part (the main body) of many PNe. The jets in PNe are thought to be propagating along the symmetry axis of the barrel-shaped region. (2) In many cases, PNe that have a general barrel morphology similar to that of the main body of SNR W49B, have one or more pairs of lobes (ears) along the direction of the jets. (3) An `H-shaped` feature is observed in many PNe. In PNe the legs of the `H-shaped' structure are thought to be parallel to the axis of the jets (e.g., Abell~63; \citealt{Mitchelletal2007}). (4) Dense filaments as observed in the SNR W49B are also observed in some PNe.
All of these morphological features can be observed in Fig. \ref{fig:1PN_elaborated} where the jet direction is marked by a light magenta arrow.
In Fig. \ref{fig:3PN_NOFEATURES} we present two other PNe, PN~G298.2-01.7 and PN~G317.1-05.7, that also have these three morphological features (barrel-shaped, lobes and `H-shaped').  
% FFFFFFFFFFFFFFFFFFFFFFFFFFFFFFFFFFFFFFFFFFFFFFFF
% FFFFFFFFFFFFFFFFFFFFFFFFFFFFFFFFFFFFFFFFFFFFFFFF
% FFFFFFFFFFFFFFFFFFFFFFFFFFFFFFFFFFFFFFFFFFFFFFFF
% FFFFFFFFFFFFFFFFFFFFFFFFFFFFFFFFFFFFFFFFFFFFFFFF

We note that in many PNe the two sides of the `H-shaped' bright feature are not symmetric to each other. Namely, the `H-shaped' region does not have a full axially symmetric morphology.  The two sides can be of unequal size and/or brightness and/or inclination. In PNe the asymmetry might result from an interaction with the ISM, or from  an asymmetrical binary interaction. For example, when a binary companion enters a common envelope it does it with a deviation from axisymmetry. But the main reason might be that the secondary star that launches the jets orbits the AGB progenitor star of the PN. If the launching period is not of many orbits, and/or the orbit is eccentric, the descendant nebula will be formed with a deviation from axisymmetry. In the case of SNR W49B the asymmetry is part of the general non-spherical explosion process of CCSNe.  

Based on analysis and observations of PNe, we take the symmetry axis of the two opposite jets to point at the two opposite lobes, and to be parallel to the legs of the ‘H-shaped’ main body. 
Two comments are in place here about the barrel shape. \cite{Mitchelletal2007} preformed kinematical analysis of the PN Abell~63 and found evidence for jets. The main body of this PN has a barrel shape, although it does not show the `bar' of an H-shape, only the two legs.
\cite{Chitaetal2008} studied the formation of bipolar nebulae formed by a rotating single blue supergiant star. They conducted numerical hydrodynamical simulations, but without jets. They obtained a bipolar structure with two lobes, but did not obtain an H-shape. These two studies further motivate us to consider jets' shaping of SNR W49B. 

\begin{figure}
\centering
\includegraphics[scale=0.75]{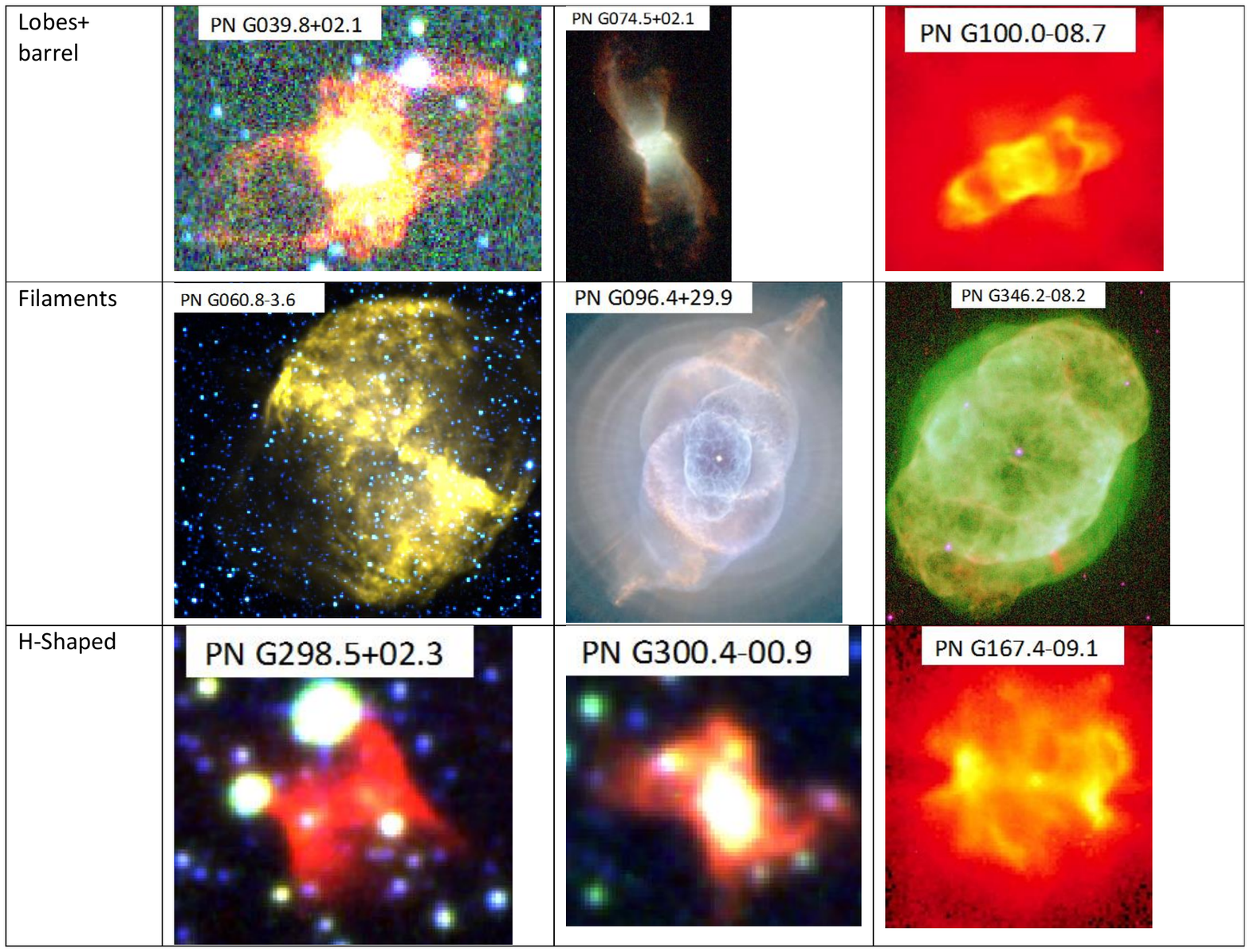}
\vskip -2.5 cm
\caption{Nine PNe that share one or more morphological features with the SNR W49B. We list the PNe according to the most prominent morphological feature, as indicated in the figure.
The PNe with their common names and credits are as follows.
PN~G 039.8+02.1 (PN K 3-17; credit: A), PN~G074.5+02.1 (NGC 6881; B), PN~G100.0-08.7 (Me 2-2; C). PN~G060.8-03.6 (M-27; A), PN~G096.4+29.9 (NGC 6543; D), PN~G346.2-08.2 (IC 4663; B), PN~G298.5+02.3; A), PN~G300.4-00.9 (Hen 2-84; A), PN~G167.4-09.1 (K 3-66; C).
Credits:
\newline
[A] HASH catalog - \citep{Parkeretal2016a,Parkeretal2016b}.
\newline
[B] HST archives from PNIC catalog - \citep{Balick1987}.
\newline
[C] \cite{SahaiTrauger1998}.
 \newline
[D] Hubbelsite from PNIC catalog \citep{Balick1987} Corradi and Tsvetanov go 9026 $http://hubblesite.org/gallery/album/nebula_collection/pr2004027a$. }
\label{fig:9PNe_SN}
 \end{figure}

We have scanned two databases of PN images, those of \cite{Parkeretal2016a,Parkeretal2016b} and of Balick (starting with \citealt{Balick1987} and extended to PNIC: Planetary Nebula Image Catalogue), and found that many PNe exhibit similar morphology features to those of the SNR W49B. We present nine such PNe in Fig. \ref{fig:9PNe_SN}. We arrange these PNe by their most prominent feature, according to three features, barrel+lobes, filaments, and `H-shaped'.
% FFFFFFFFFFFFFFFFFFFFFFFFFFFFFFFFFFFFFFFFFFFFFFFF
% FFFFFFFFFFFFFFFFFFFFFFFFFFFFFFFFFFFFFFFFFFFFFFFF

\begin{figure}
\centering
\hskip -4.5 cm
\includegraphics[scale=1.0]{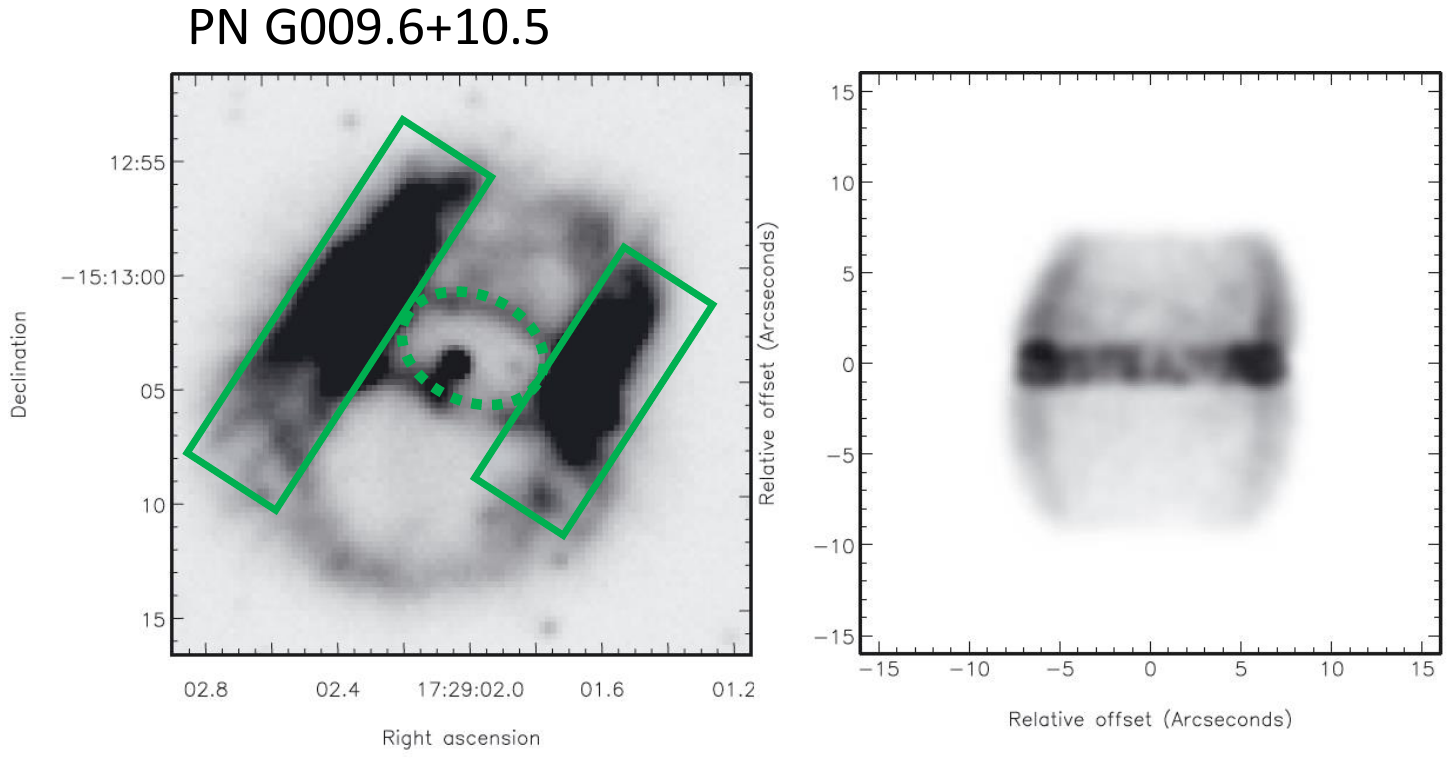}
\vskip -15.5 cm
\caption{The PN Abell~41 (PN~G009.6+10.5). The left panel is an [NII]~6584$\text{\AA}$ image. The right panel presents a 
synthetic model for the emission in this line from Abell~41 displayed
at an inclination of $90^\circ$, namely, the symmetry axis is on the plane of the sky and vertical in the image. Both panels are taken from  \cite{Jonesetal2010}. On the left panel we added marks of the ring and the two legs of the `H-shaped' region. }
\label{fig:Abell41}
 \end{figure}

The line connecting the two `H-legs' in PNe is considered to be an equatorial ejected material rather than a jet. It can be a ring (or torus) viewed edge-on, or a segment of a ring. We suggest the same for the CCSNR W49B. One good case of such a PN is Abell~41 (PN~G009.6+10.5) that was analyzed by \cite{Jonesetal2010}. We present both their observation and analysis in Fig. \ref{fig:Abell41}; in the left panel we present the [NII]~6584$\text{\AA}$ image and in the right panel their synthetic model. A clear equatorial ring is observed. \cite{Jonesetal2010} analyzed the structure of Abell~41, and built a synthetic model for its structure. When viewed edge-on (inclination angle of $90^\circ$), a dense bar is seen extending perpendicular to the symmetry axis. In PNe the symmetry axis is along the axis of the jets. Namely the dense bar is perpendicular to the jets.    
% FFFFFFFFFFFFFFFFFFFFFFFFFFFFFFFFFFFFFFFFFFFFFFFF
% FFFFFFFFFFFFFFFFFFFFFFFFFFFFFFFFFFFFFFFFFFFFFFFF

Based on the strong similarities between the morphology of SNR W49B and many PNe, we suggest that this SNR was shaped by two opposite jets launched from the center during, or shortly after, the explosion.
We further suggest that, like in PNe, the symmetry axis of the jets is along the symmetry axis of the barrel-shaped inner region, namely, parallel to the legs of the `H-shaped' region seen in different X-ray lines. Our proposed axis of the jets is perpendicular to the jets' axis as proposed by \cite{Lopezetal2013} and \cite{Keohaneetal2007}. We take the structure these papers refer to as jets, to be an equatorial dense gas, as seen in many PNe. 

Although our conclusion is based on a qualitative comparison of the CCSNR W49B with PNe, we feel it stands on a solid ground. There are similarities between bipolar structures carved by jets in PNe and many other astrophysical objects, such as X-ray deficient bubbles in cooling flow clusters \citep{SokerBisker2006}, ears in SNR of Type Ia SNe \citep{TsebrenkoSoker4502015}, and nebulae of massive stars, e.g., Eta Carinae (e.g., \citealt{Morseetal1998}). As these and other papers show, the similarities go much beyond the morphological similarities, and extend to physical processes, including the interaction of the jets with the ambient gas. The orientation of the jets in SNR W49B has implications on the explosion mechanism of the SN. In particular, our suggested jets' orientation hints that the jets played an important role in the explosion, as we argue next.

% ==========================================================
\section{THE ENERGY OF THE JETS}
\label{sec:jets}
\begin{figure}
\centering
\includegraphics[scale=0.65]{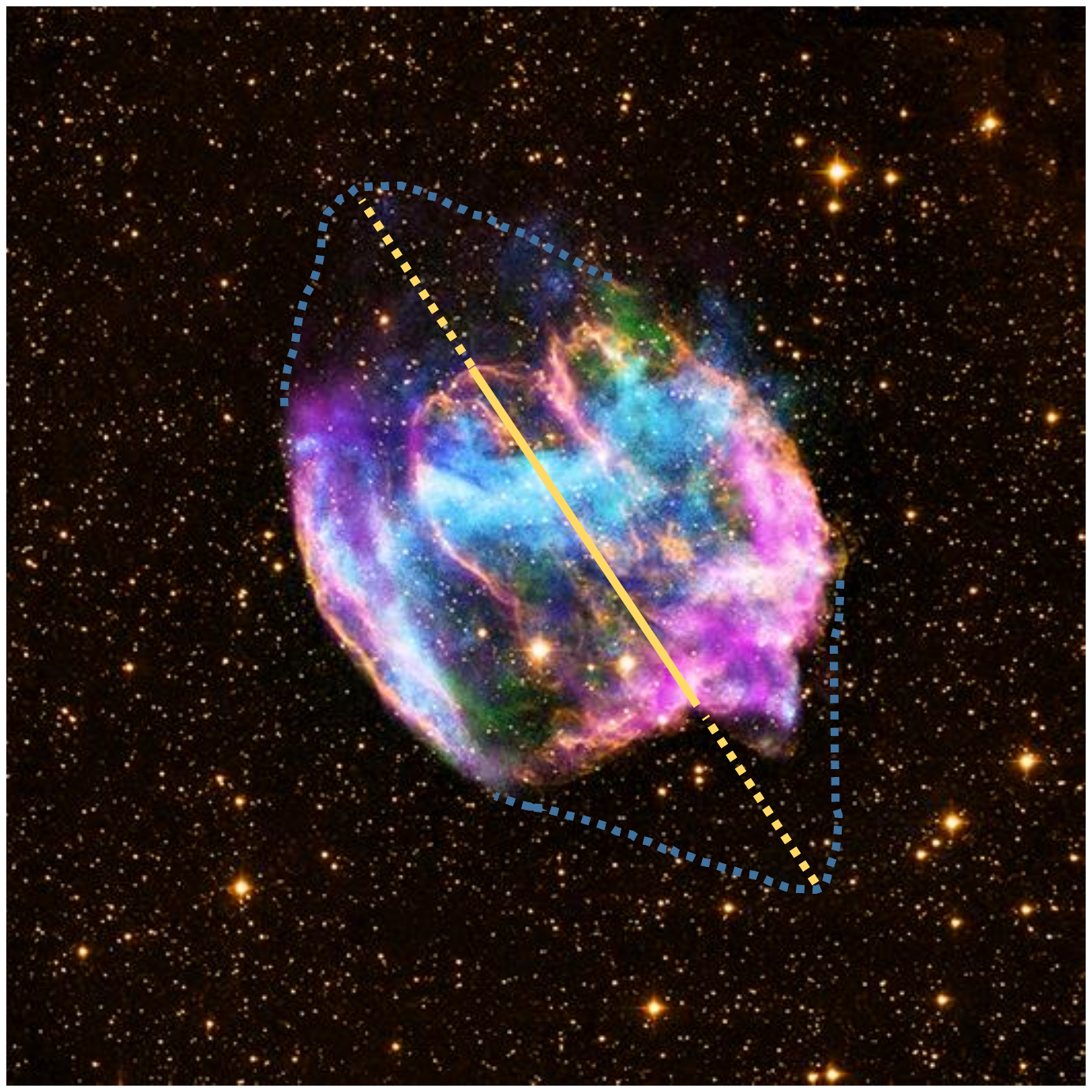}
\vskip -5.5 cm
\caption{ The figure shows the CCSNR W49B with the suggested symmetry axis marked with a yellow line. The suggested lobes are marked by their approximate hypothetical boundaries with dotted cyan line. The suggested lobes have a diameter marked with a dotted line, the length of which is equal to half the length of the solid yellow line. Credit: Chandra website  (http://chandra.harvard.edu/photo/2013/w49b/).}
\label{fig:Lobes_SN}
 \end{figure}

% ==========================================================
In Fig. \ref{fig:Lobes_SN} we mark the suggested lobes that were inflated by the jets we are claiming for in the study. We suggest these lobes based on comparison to PNe. When viewing the PNe (see figure \ref{fig:9PNe_SN}) we can generally say that the lobes diameter is of the same order as the barrel radius.
If we extrapolate this notion to the SNR W49B we can suggest that with deeper observations, possibly at different wavelengths also, the lobes might be revealed.
% FFFFFFFFFFFFFFFFFFFFFFFFFFFFFFFFFFFFFFFFFFFFFFFF
% FFFFFFFFFFFFFFFFFFFFFFFFFFFFFFFFFFFFFFFFFFFFFFFF

Our crude estimate is based on the structure of many PNe where the distance of the tip of the lobe from the center is twice the distance of the edge of the barrel-shaped region from the center (each dashed yellow line is half the length of the solid yellow line). Nonetheless, it gives us the possibility to estimate the lobes energy as we anticipate based on the morphologies of many PNe.

 We follow \cite{GrichenerSoker2017} and estimate the energy of the suggested lobes with the same assumptions they have made (we will not repeat the description of the technique here). We find that the combined kinetic energy of the two assumed lobes is about $20-25 \%$ of the energy of the entire SNR shell. Since not all the energy of the jets will end up in the lobes, we crudely estimate that the jets that inflated the lobes and shaped the CCSNR into a barrel shape had a kinetic energy that amounts to about one quarter to one third of the energy of the entire CCSNR.

This energy is more than twice as large as the energy of the jets in the eight CCSNRs studied by \cite{GrichenerSoker2017}. It seems that the jets were strong enough to disperse the lobes/ears of SNR W49B to the degree that they are too faint to be detected. The lobes might be larger than what we assumed in drawing Fig. \ref{fig:Lobes_SN}, and therefore, the energy of the jets might be larger even than the value we deduce here.

% ==========================================================
\section{SUMMARY}
\label{sec:Summary}
% ==========================================================

We compared the morphology of the CCSNR W49B with those of many PNe, 13 of which are presented here in Figs. \ref{fig:1PN_elaborated} - \ref{fig:Abell41}. Although this comparison is qualitative, we consider the following conclusions to be robust and with important implications.
\newline
(1) Based on the similar structure of the barrel-shaped main inner part of SNR W49B with many PNe, we deduced that the symmetry axis of SNR W49B is along the symmetry axis of the `barrel', as can be seen in Figs. \ref{fig:1PN_elaborated}, \ref{fig:3PN_NOFEATURES} and \ref{fig:Lobes_SN}.
\newline
(2) The shaping mechanism of the PNe with similar structures is thought to be jets. We deduced that SNR W49B was shaped by two opposite jets launched along the symmetry axis of the `barrel'. This direction is perpendicular to earlier suggestions for the jets' direction (e.g., \citealt{Keohaneetal2007, Lopezetal2013}).
\newline
(3) In most PNe two opposite lobes/ears protrude along the symmetry axis. We assumed that such lobes/ears exist in SNR W49B, but that they were dispersed and hence too faint to be observed. We very crudely (and conservatively) estimated that the energy of the jets that inflated these lobes/ears was about one quarter to one third of the energy of the entire CCSNR.

One of the expectations of the JFM explosion mechanism of CCSNe is that the last segments of the jets to be launched will expand to large distances. The reason is as follows (see also \citealt{GrichenerSoker2017}). According to the JFM, as long as accretion process continues, jets are launched (assuming their is enough angular momentum to form an accretion disk or an accretion belt; \citealt{SchreierSoker2016}). The accretion process continues as long as there is bound core material, as is required for a negative feedback mechanism.
After the outer parts of the core are completely ejected some in-flowing gas, that was already on its way in, is accreted and the final jet-launching episode takes place. The gas in this last episode does not encounter dense gas in the core and envelope anymore, as the core and envelope have been ejected. The jets expand freely and interact with the exploding material outside the star, during the expansion phase. Such jets can form ears \citep{TsebrenkoSoker2013}.

The similarity of the morphology of SNR W49B with those of many PNe further support the call for a paradigm shift from neutrino-driven to jet-driven core-collapse supernova mechanisms.

We thank an anonymous referee for detailed and useful comments. This research was supported by the Israel Science foundation. This research has made use of the HASH PN database at hashpn.space.

\end{document}